\documentclass{article}

\usepackage{arxiv}
\makeatletter
\def\@seccntformat#1{\@ifundefined{#1@cntformat}%
   {\csname the#1\endcsname\quad}  
   {\csname #1@cntformat\endcsname}
}
\let\oldappendix\appendix 
\renewcommand\appendix{%
    \oldappendix
    \newcommand{\section@cntformat}{\appendixname~\thesection\quad}
}
\makeatother
\usepackage{natbib}
\usepackage{amsmath,amssymb}
\usepackage[utf8]{inputenc} 
\usepackage[T1]{fontenc}    
\usepackage{hyperref}       
\usepackage{url}            
\usepackage{booktabs}       
\usepackage{amsfonts}       
\usepackage{nicefrac}       
\usepackage{microtype}      
\usepackage{lipsum}
\usepackage{caption}
\usepackage{subcaption}
\usepackage{xspace}
\usepackage{graphicx}
\usepackage{floatrow}   

\usepackage[framemethod=tikz]{mdframed}

\title{Temporal dynamic model for resting state fmri data: A Neural Ordinary Differential Equation approach}

\author{
  Zheyu Wen\\
  Department of Electrical Engineering and Computer Science\\
  University of Michigan\\
  Ann Arbor, MI 48105 \\
  \texttt{zheyw@umich.edu} \\
}

\begin{document}
\maketitle

\begin{abstract}
The objective of this paper is to provide a temporal dynamic model for resting state functional Magnetic Resonance Imaging (fMRI) trajectory to predict future brain images based on the given sequence. To this end, we came up with the model that takes advantage of representation learning and Neural Ordinary Differential Equation (Neural ODE) to compress the fMRI image data into latent representation and learn to predict the trajectory following differential equation. Latent space was analyzed by Gaussian Mixture Model. The learned fMRI trajectory embedding can be used to explain the variance of the trajectory and predict human traits for each subject. This method achieves average 0.5 spatial correlation for the whole predicted trajectory, and provide trained ODE parameter for further analysis.
\end{abstract}


\section{Introduction}
Figuring out spatial temporal relationship in functional Magnetic Resonance Imaging(fMRI) trajectory is a grand challenging problem. The field lacks a explainable and accurate model to fit the measured data. We are going to provide a new method to fit the fMRI trajectory. There are two main challenges in the field. First, we don't know what's the process of resting state fMRI data trajectory(\cite{liegeois2017interpreting}). Second, it's difficult to eliminate random noise and physiological noise in data, which makes the patterns more complex (\cite{laumann2017stability}). AutoRegressive model(\cite{zalesky2014time}) and Hidden Markov Model( \cite{vidaurre2017brain}), the two well known methods for explaining temporal dynamic, are far from understanding the process well.

Many work are dedicated to analyzing the spatial temporal dynamic regarding the measured fMRI data (\cite{lurie2020questions},\cite{chang2010time}\cite{liegeois2017interpreting}). Some intrinsic properties are exploited from the observation data without an explicit mathematical model to interpret. No explainable and accurate model is proposed except some models based on Region Of Interest(ROI) with linear/non-linear Gaussian hypothesis (\cite{liegeois2017interpreting}). Those works don't provide a exploration of latent representation of measured fMRI data for spatial temporal information, which can be utilized to model the whole brain data. Neural network can help explore the principle of transition between spatial temporal representation with a predefined network architecture. Our work provide a interpretable model that can help predict future brain map or what happened between two given trajectories to interpolate for original data. The main idea is to compress the trajectory into spatial temporal latent representation and use a backbone of video prediction model to constrain the representation and help do forward prediction. In this work, we are not going to give physiological interpretation of the temporal relationship of measured fMRI data but to provide a new way of spatial temporal modeling on fMRI trajectory.

Predicting what fMRI data will be in the future can be viewed as a video prediction problem. Video prediction is challenging because of its uncertainty (\cite{jayaraman2018time}). Network that predict video frames recursively will accumulate blurry in images which makes the prediction unusable after several time points. To mitigate this problem and make the prediction with high quality, representation learning can be used to compress the spatial temporal information in the bottleneck and temporal dynamic model can take advantage of these representation to do forward prediction. Our contribution is that we propose to use Neural Ordinary Differential Equation (Neural ODE) as video prediction backbone combined with a spatial temporal representation learning scheme to learn latent information of a group of fMRI images. 

The paper is organized as follows, in section 2, we compare our model with other representation learning and video prediction model. We will also compare the difference between traditional temporal dynamic model for resting state fMRI data with ours. In section 3, formulation of Neural ODE will be introduced and we will explore the usage of learned spatial temporal representation. In section 4, we specify the experimental setting and result. In section 5, we discuss the key factors of success in our model and the potential usage of it in temporal relationship exploration for fMRI data.

\section{Related work}
Time varying functional connectivity is a popular terminology in describing temporal relationship of fMRI data discussed in \cite{lurie2020questions} and \cite{liegeois2017interpreting}. There are two popular methods in temporal dynmaic modeling for resting state fMRI data which are Auto Regressive (AR) model and Hidden Markov Model(HMM). \cite{liegeois2017interpreting} and \cite{zalesky2014time} used AR model as a simple linear model to fit the trajectory of ROI of resting state data and generate null data to test whether dynamic functional connectivity is significant enough. HMM is used to find the hidden state underneath the data. \cite{vidaurre2017brain} found there are two meta-states among tens of state got from HMM. However these methods are all based on ROI which can not recover the whole brain activity after establishing the temporal model. Previous study also introduce PCA and ICA to compress the whole data into a 1D array and analyze based on this. \cite{kim2020representation} introduced beta Variational AutoEncoder to compress the transformed 2D image into 1D array while disentangling the latent factors into different latent variables. Inspired from this, we develop a method to compute the spatial temporal representation to model the whole trajectory for each subject. Other methods used to explore functional connectivity dynamic are introduced in work of \cite{cabral2017functional}, \cite{kashyap2019dynamic}, \cite{chang2010time} and \cite{laumann2017stability}.

In our setting, we are going to predict future fMRI images based on the given sequence, which can be viewed as a video prediction problem. This problem has two main categories, which are forward prediction and bidirectional prediction. Making future prediction while maintaining high quality of image is hard. There are mainly two ways to deal with this problem. The first method in \cite{oh2015action} learn the prediction in the image space. While \cite{ranzato2014video} and  \cite{jayaraman2015learning} learned the temporal dynamic information of video frame in latent space. Recently time agnostic video prediction proposed by \cite{jayaraman2018time} lend new view point to the field. In our model, we train the model to learn spatial temporal latent representation for a batch of images while adapting Neural ODE to fit the learned representation.

Neural ODE was first introduced in \cite{chen2018neural}. Later on more machine learning method related with ODE appear, including Augmented ODE introduced in \cite{dupont2019augmented}, second order ODE in \cite{yildiz2019ode2vae}, and Stochastic Differential Equation (SDE) in \cite{jia2019neural} provide more methods to model the temporal relationship of given data. Here we adapt Augmented ODE in our experiment that enlarge the representation space of Neural ODE, which is beneficial for us to improve the quality of prediction and makes it possible to predict beyond the space of input fMRI data trajectory to account for the variation of complex data. Another method in using Neural Network to fit fMRI data was proposed in \cite{kashyap2020brain} 
\cite{khazaee2017classification} and \cite{du2018classification} explored fMRI data temporal relationship which is useful for disease diagnosis  or providing insight into fMRI data trajectory as shown in  \cite{tagliazucchi2012criticality}. In our work, we use the learned spatial temporal latent representation to do the human traits prediction. This method can also be useful in individual classification and critical point analysis.

\section{Spatial Temporal modeling on rsfMRI data}
Our goal is to fit observation rsfMRI data into Neural ODE Network to make it possible to predict the future fMRI data given the input trajectory. We can model it as a video prediction problem. In video prediction, the goal is to predict the future frames given the first several input frames. There are two basic tasks. The first one is to do the forward prediction, and the target is the all future frames. The second task is the bidirectional prediction. We are given the first and the last several frames, and the target is to do the interpolation. In this section, we will talk about how to train the Ordinary Differential Equation using neural network to fit resting state fMRI data trajectory. We will also explain how the variance of the future prediction is accounted in latent representation of trajectory. The latent representation will contribute to some downstream analysis.
\subsection{Basic definition}
Human Connectome Project (HCP) provides resting state fMRI data records for different subjects. Each subject provides 1200 time points volumetric data with data size $91\times109\times91$. Each Volumetric data corresponding to a frame in video prediction context. We will adapt the dimension reduction method used in \cite{kim2020representation} to reduce the Volumetric data dimension into 2D data with size $192\times 192$. We denote the $j_{th}$ 2D fMRI data frame of subject $i$ as $X_{i,j}$. Our aim is to use the given first $j$ video frames $X_{i, 0:j-1}$ to predict the future frames of subject $i$, which is $X_{i, j:J}$, where $J$ is maximum number of time point in record.
\subsection{Neural ordinary differential equation}
 Video prediction is a challenging problem since the prediction quality may drop dramatically after several time points prediction. The uncertainty in video frames also contribute to the difficulty of predicting accurate future frames. Here we first try to solve the problem of prediction quality by using latent representation and Neural Ordinary Differential Equation introduced in \cite{dupont2019augmented}. Uncertainty will be explained by Variational AutoEncoder in next subsection.\\
\begin{figure}[H]
    \centering
    \begin{subfigure}[b]{0.5\textwidth}
        \centering
        \includegraphics[width=\textwidth]{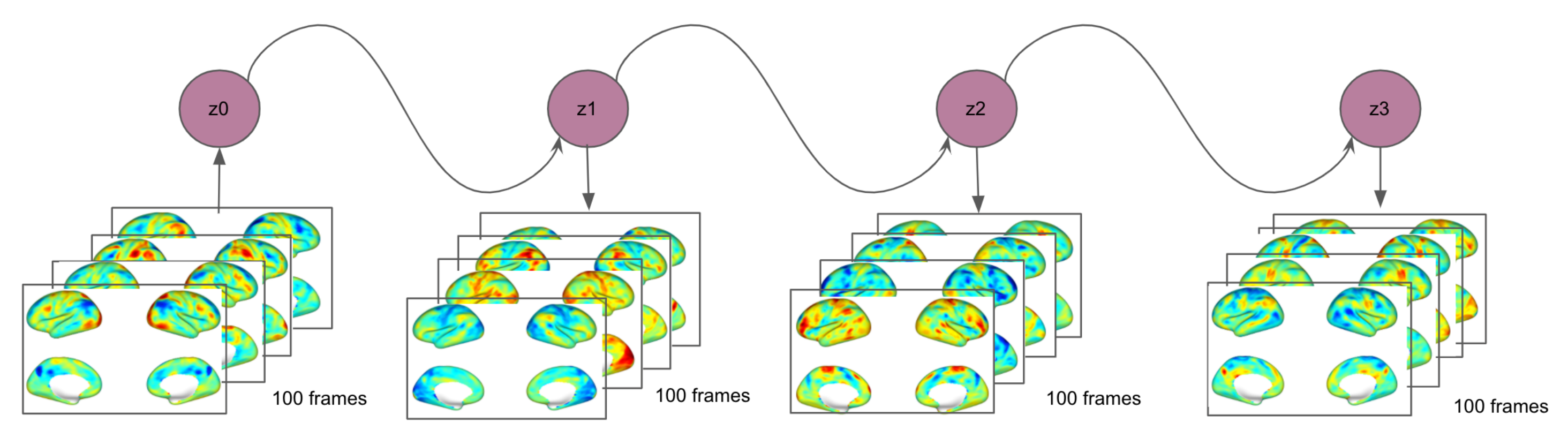}
        \caption{Frames in one fMRI trajectory are grouped into four sets. Encoded into spatial temporal latent representation and fitting ODE for prediction. Predicted latent code is decoded to image space.}
        \label{fig:general_archi}
    \end{subfigure}
    \begin{subfigure}[b]{0.45\textwidth}
        \centering
        \includegraphics[width=\textwidth]{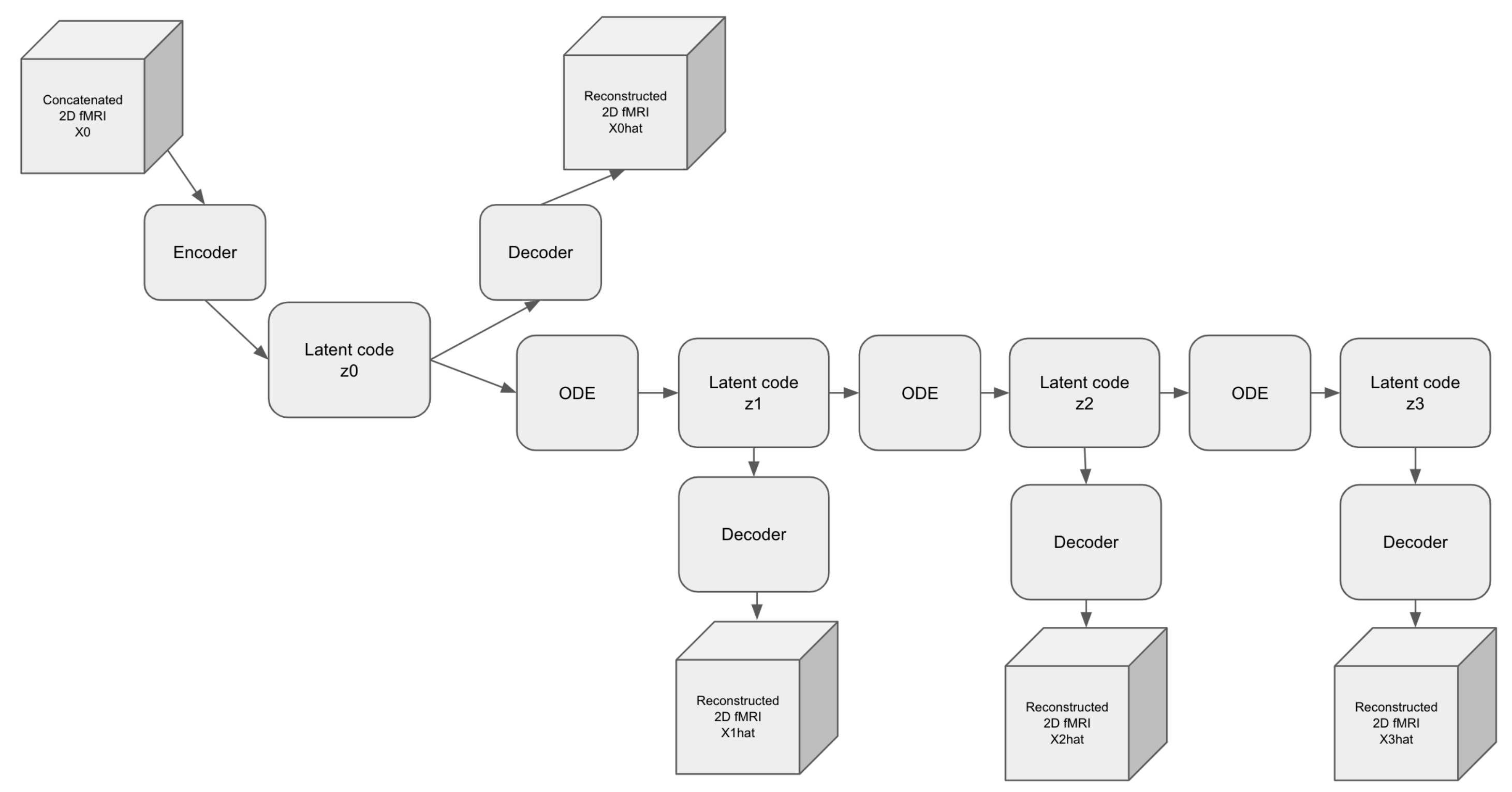}
        \caption{Spatial Temporal latent representation learning with Ordinary Differential Equation as backbone to explore temporal dynamic of original data and predict future frames.}
        \label{fig:RNN_train}
    \end{subfigure}
    \caption{Video generation scheme learning the spatial temporal latent representation from groups of fMRI images and do forward prediction with Oridinary Differential Equation as backbone.}
    \label{fig:ODE_model}
\end{figure}
Since resting state fMRI data changes very slowly during recording time, we can first do the downsampling of the original 1200 time points into 400 time points to reduce the prediction length while not influencing the utility of our result. Moreover, we do a innovative data processing to group the whole fMRI data trajectory into 4 groups for each individual, in which each group has 100 data frames. As shown in Fig. \ref{fig:general_archi}, we first use encoder to compress the 3D concatenated data (the first dimension is time while the last two dimensions are height and width of the 2D image frame) into a 1D array with length 64 to represent the spatial temporal information for fMRI sequence in 100 time points, which is denoted as $z_0$. The reason of selecting 100 as the group size is that the length of the data can successfully reveal the spatial temporal information for a certain subject. The encoder introduced here can be written as
\begin{equation}
    \label{eqn:encode}
    z_0 = f_{\theta}(X_{i, 0:99})
\end{equation}
We will not include subscripts for $z_t$ to avoid mass in our notation but $z_0$ is determined by input data for a certain subject $i$. $f_{\theta}$ is encoder and $\theta$ is parameter of encoder network. Then we apply Neural ODE on the latent representation to forward predict $z_t (t\in \mathbb{N}_{+})$ . The differential equation are shown as following
\begin{equation}
    \label{eqn:ODE function}
    \frac{dz_t}{dt} = W_2\Phi(W_1[z_t, \mathbf{0}] + b_1) + b_2
\end{equation}
in which $W_1$, $W_2$, $b_1$, $b_2$ is linear network parameter and $\Phi$ is nonlinear function. Here we concatenate $\mathbf{0}$ with $z_t$ to enlarge its representation space following method of Augmented Neural ODE (\cite{dupont2019augmented}). By solving the differential equation, we can obtain the latent representation in the future by following
\begin{equation}
    \label{eqn:integral on ODE}
    \hat{z}_{t_1 + \Delta t} = z_{t_1} + \int_{t_1}^{t_1 + \Delta t} \frac{dz_t}{dt}|_{t=t_1} dt
\end{equation}
Lastly, we decode the predicted spatial temporal representation to recover the data in image space following
\begin{equation}
    \label{eqn:decode to image space}
    \hat{X}_{i, t*100:(t+1)*100-1} = g_{\phi}(z_t)
\end{equation}
where $g$ is the decoder and $\phi$ is parameter of it. Remind that we have four groups of data for each individual, the first group is used as input while the left three is regarded as output. The loss function is written as following
\begin{equation}
    \label{eqn:loss for forward pred}
    Loss = \sum_{i=1}^b\sum_{t=0}^3||\hat{X}_{i, t*100:(t+1)*100-1} - {X}_{i, t*100:(t+1)*100-1}||_2
\end{equation}
where $b$ is batchsize in the training and ${X}_{i, t*100:(t+1)*100-1}$ is ground truth of the estimated fMRI data trajectory. We use Mean Squared Error as training loss between decoded images with ground truth images. 

\textbf{spatial correlation:}
In testing time, we do forward prediction given the first group of images. Spatial correlation is used to evaluate the performance of trained network
\begin{equation}
     \label{eqn:spatial correlation}
    R(X, \hat{X}) = \frac{C_{x, \hat{x}}}{\sqrt{C_{x,x}C_{\hat{x}, \hat{x}}}}
\end{equation}
where $C$ is covariance matrix. $x$, $\hat{x}$ are vectorized form of $X$ and $\hat{X}$. The high spatial correlation value means the similarity between the estimated images and ground truth images.

\textbf{bidirectional prediction:}
While the above description focus on forward prediction, it is easy to generalize to bidirectional prediction. In this situation, for each subject, we only know the first 200 data and last 100 data in training. We can assign latent representation to the trajectory it belongs to so the three predicted latent representation will be decoded to three known groups of images. Neural ODE can interpolate the left 100 time point for us while fitting the data on both end of trajectory. The loss function will be tweaked as
\begin{equation}
    \label{eqn:loss for bidirectional pred}
    Loss = \sum_{i=1}^b\sum_{t=0,1,3}||\hat{X}_{i, t*100:(t+1)*100-1} - {X}_{i, t*100:(t+1)*100-1}||_2
\end{equation}
Therefore, we use the spatial temporal latent representation for different groups of frames. The network do forward predictions or bidirectional prediction of frames by first predicting on latent representation and then decode to image space.\\
\subsection{Variance explanation introduced by VAE}
To further explain the variance in spatial temporal latent representation, we introduce the Variational AutoEncoder (VAE) following \cite{kingma2013auto} here to model the latent space as prior Gaussian distribution. In previous subsection, we use AutoEncoder(AE) to encode and decode the fMRI data latent representation. Here we make a slight change of this part. The VAE encoder encodes the group of images into a latent distribution. We then adapt reparametrization tricks in \cite{kingma2013auto} to sample from this distribution and get latent representation. The distribution established here explains the variance of latent code for different subjects. The encoder part will be changed into 
\begin{equation}
    \label{eqn:vae distribution}
    [\mu_0, logvar_0] = f_{\theta}(X_{i, 0:99})
\end{equation}
\begin{equation}
    \label{eqn:reparametrize}
    z_0 = reparam(\mu_0, logvar_0)
\end{equation}
where $\mu_0$ and $logvar_0$ are mean and log variance of spatial temporal latent distribution, and $z_0$ has the same meaning as in previous subsection, which is spatial temporal representation of the given trajectory. Since we model the prior distribution of latent space as white Gaussian, here we use KL divergence to minimize the loss between approximate posterior distribution $p(z_0|X_{i, 0:99})$ and given prior $\mathcal{N}(0, 1)$.
\begin{equation}
    \label{eqn: loss for vae}
    Loss = \sum_{i=1}^b\sum_{t=0}^3||\hat{X}_{i, t*100:(t+1)*100-1} - {X}_{i, t*100:(t+1)*100-1}||_2 + D_{KL}(q(z|X_{i,0:99}), p(z))
\end{equation}
in which $p(z)$ is prior Gaussian.

\subsection{Other temporal dynamic model}
In the model mentioned above, we use ODE as backbone to predict spatial temporal latent representation $z_t$ given $z_0$. Here we will try a different backbone and a pure RNN model without spatial temporal representation learning. For model shown in Figure \ref{fig:ODE_model}, we replace ODE by Recurrent Neural Network(RNN) as shown in Figure \ref{fig:rnn_w_ae_archi}. RNN take $z_{t-1}$ as input and output $z_t$ with other setting the same. We also want to test effectiveness of the latent representation $z_t$, so we tried a intuitive RNN architecture shown in Figure \ref{fig:rnn_wo_ae_archi}. We called it pure RNN model, in which the RNN take fMRI 2D image in one time point as input and output the image in the next time point. So in pure RNN we do one time point forward prediction rather than predicting a group of fMRI images.

\subsection{Gaussian mixture model analyze latent space}
We model the prior distribution of $z_0$ as Gaussian distribution, while in Neural ODE, $z_t$ will not necessarily follows the Gaussian distribution. It's hard to use one Gaussian to explain the latent space. So here we introduce Gaussian Mixture Model (GMM) as shown in \cite{reynolds2009gaussian} to help explain and do clustering to see the common patterns among all the groups of images. We model the whole latent space as a Gaussian Mixture.
\begin{equation}
    \label{eqn:gmm basic model}
    p(z_t) = \sum_{k=1}^K\pi_k\mathcal{N}(z_t|\mu_k, \Sigma_k)
\end{equation}
where $t=0,1,2,3$, and $\mu_k$, $\Sigma_k$ are the mean and variance of each Gaussian distribution, and $\pi_k$ is weight for this Gaussian distribution. We will use this model to do clustering on latent space to see how many Gaussian distributions' mixture can best explain the latent space. The process of GMM clustering is as follows. For a specific latent code $z_{t}$ of time point $t$. Suppose probability of data $z_{t}$ in cluster $k$ $(k=1,...,K)$ is $\gamma(s_{tk})=p(s_{tk}=1|z_t)$. The log probability of data $z_t$ is 
\begin{align}
    \label{eqn:gmm max prior dist}
    \ln{p(z_t)} &=\sum_{k=1}^K \gamma(s_{tk})\ln{p(z_t)}  \\
     &= \sum_{k=1}^K \gamma(s_{tk})\ln{\frac{p(z_t, s_{tk}=1)}{p(s_{tk}=1|z_t)}} \\
     &= \sum_{k=1}^K \gamma(s_{tk})\ln{p(z_t, s_{tk}=1}) - \sum_{k=1}^K \gamma(s_{tk})\ln \gamma(s_{tk})
\end{align}
We maximize the log probability of $z_t$ to iteratively update variable $\gamma(s_{tk})$, $\pi_k$, $\mu_k$ and $\Sigma_k$. EM algorithm is used here to solve this cluster problem. For more information of EM algorithm in our setting, see Appendix \ref{Appendix B}. 
\subsection{Gradient flow analysis on ODE}
Ordinary Differential Equation reveals gradient magnitude and direction of $z_t$ when given current $z_t$ value. We care about how the system looks like before and after training. We plot the gradient flow of first two latent variable in $z_t$ for random initialized ODE model and there are several equilibrium in the whole system. On the sub-diagonal, the gradient flow run fast towards negative value for both first two latent variables in latent code. The initialized system may not be stable and as prediction goes, latent code may diverge fast.
\begin{figure}[H]
    \centering
    \includegraphics[width=0.3\textwidth]{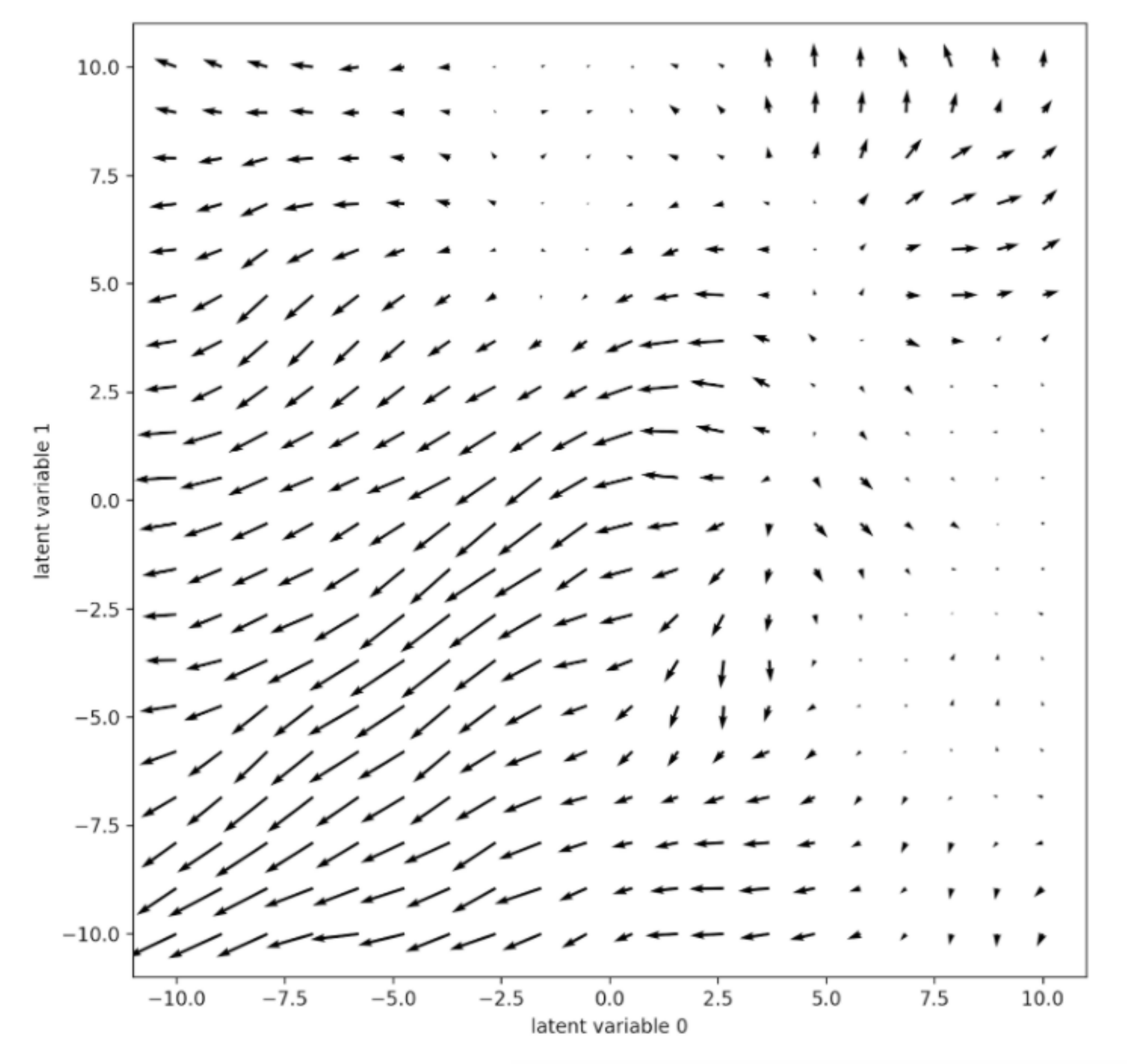}
    \caption{gradient flow for random initialize ODE.}
    \label{fig:randomized_gradient}
\end{figure}
We can compute the equilibrium by setting Equation \ref{eqn:ODE function} to 0 and got a equilibrium set $\{z_t|\frac{dz_t}{dt}=0\}$. Latent code around equilibrium will also be analyzed and corresponding temporal correlation for decoded estimation fMRI trajectory will be computed as Pearson product-moment correlation coefficients.
\begin{equation}
    \label{eqn:temporal correlation}
    R(x,y) = \frac{\sum(I(x)_i-\Bar{I}(x))(I(y)_i-\Bar{I}(y))}{\sqrt{(\sum(I(x)_i-\Bar{I}(x))^2)(\sum(I(y)_i-\Bar{I}(y))^2)}}
\end{equation}
where $I(x)_i$ is pixel value at coordinate $x$ for time point $i$ while $\Bar{I}(x)$ is mean of pixel value in $x$ among all the time points. Let's assume $x$ is the seed we select and fix. $y$ is varied to allow us compute temporal correlation for all the pixel with the seed in image.
\section{Experiment}
We have proposed a new way of video prediction in resting state fMRI data. We focused our evaluation on quality of frame generation and explanation of spatial temporal latent code. Two other architecture were implemented to be compared with. The architectures are available in appendix A. There are mainly three downstream tasks. First we used GMM to explain the geometry of spatial temporal latent space. Second we tried to see how temporal correlation changes along the given gradient flow provided by trained Neural ODE. Thirdly, we concerned how to use latent code to predict human traits of subjects in HCP data.
\subsection{data preparation and model architecture}
The data was collected from HCP 3T resting state fMRI data. We applied detrend, band pass filter and standardize the signal, then extracted grey matter of brain and apply the transformer following \cite{kim2020representation} to convert Volumetric data into 2D fMRI data. For single volumetric fMRI data, two 2D images were obtained for left half sphere and right half sphere separately. Three hundreds subjects were used to train the network, fifty subjects were used in validation and one hundred fifty subjects were used for testing. \\

Our target is to extract spatial temporal latent code for fMRI data. We had a spatial encoder for both left and right half sphere 2D data to extract a latent representation for spatial information. We first added a dimension of channel for these 2D pre-processed data and input to Conv2D layer with output channel 32, kernel size 8, stride 2 and padding 0, then we concatenated data from left sphere and right sphere on the channel dimension. The concatenated data went through four Conv2D layers with output channel size 128, 128, 256, 256, each layer had kernel size 4, stride 2 and padding 0. Then we resized the output of last layer to 1D array and used one linear layer to transform into a vector of length 256. This is the spatial latent code for each data frame. We then concatenated spatial latent code of 100 consecutive frames into a matrix with size $100\times 256$. We further used 2 Conv2D layers with output channel 2, 4 separately to encode this spatial temporal code matrix. Each layer has kernel size 3, stride 2 and padding 1. These two layers extract the spatial temporal information in this latent representation matrix and go through a linear layer to output $z_0$ with length 64. ODE network follows Equation \ref{eqn:ODE function} with $tanh$ as nonlinear function. $\texttt{torchdiffeq}$ package was used to do forward and backward propagation for training Neural ODE.

The decoder architecture is symmetric to the encoder. We first did linear transformation to spatial temporal latent code and feed into two Convolutional transpose 2D layer with output channel size 2, 1, and padding 0, 1 seperately. The kernel size is 3 and stride maintains 2. Then feed the spatial latent information into four 2D Convolutional transpose layers with padding changed to 3. Lastly the output is chunked into two part in its channel dimension and feed into another 2D Convolutional transpose layers with padding 1. Other setting in these two layers are symmetric with encoder. 
\subsection{Training and spatial correlation}

\begin{figure}[H]
    \centering
    \begin{subfigure}[b]{0.3\textwidth}
        \centering
        \includegraphics[width=\textwidth]{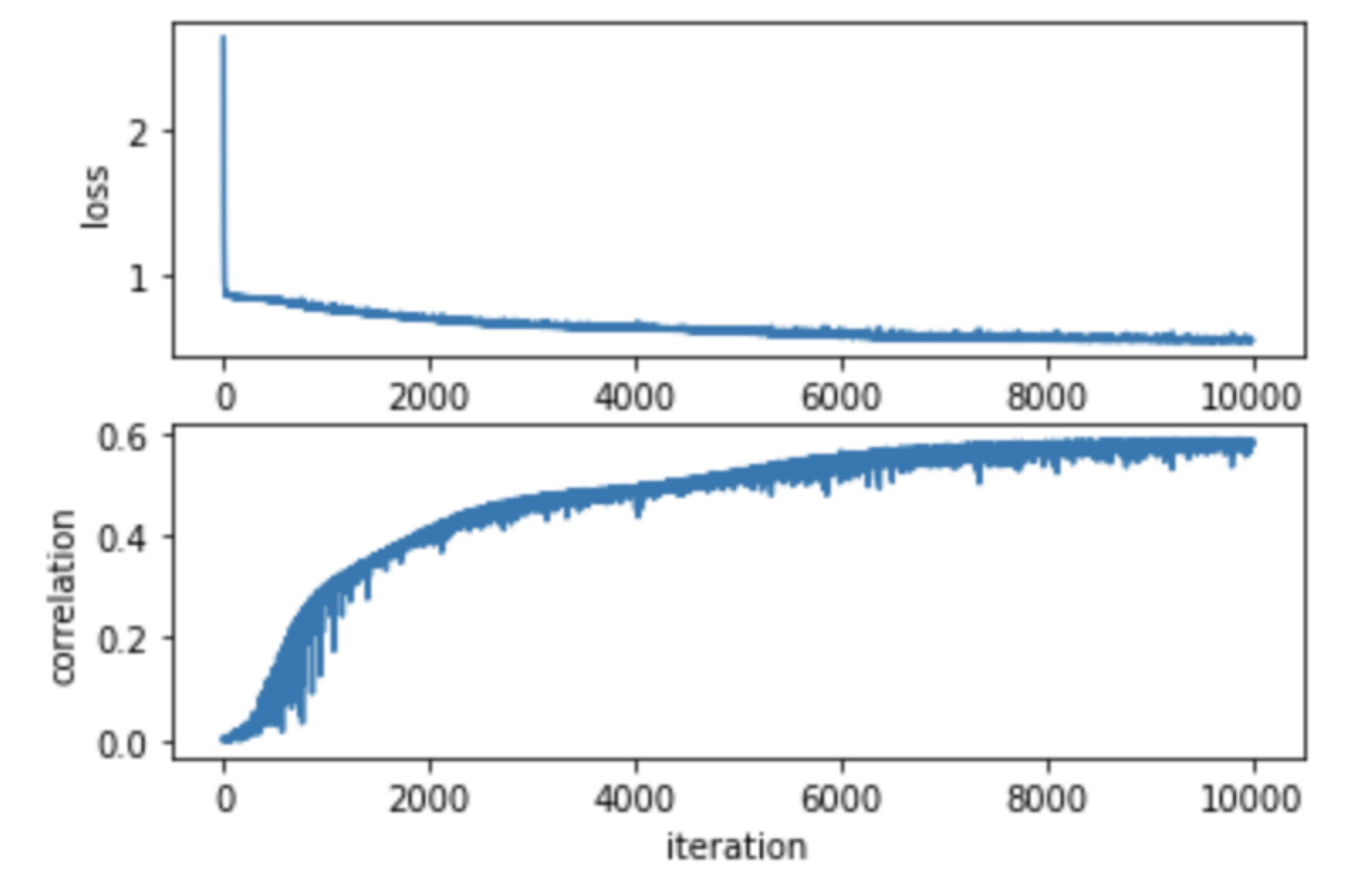}
        \caption{ODE model loss curve in training(upper panel), spatial correlation(lower panel)}
        \label{fig:ODE_train}
    \end{subfigure}
    \hfill
    \begin{subfigure}[b]{0.3\textwidth}
        \centering
        \includegraphics[width=\textwidth]{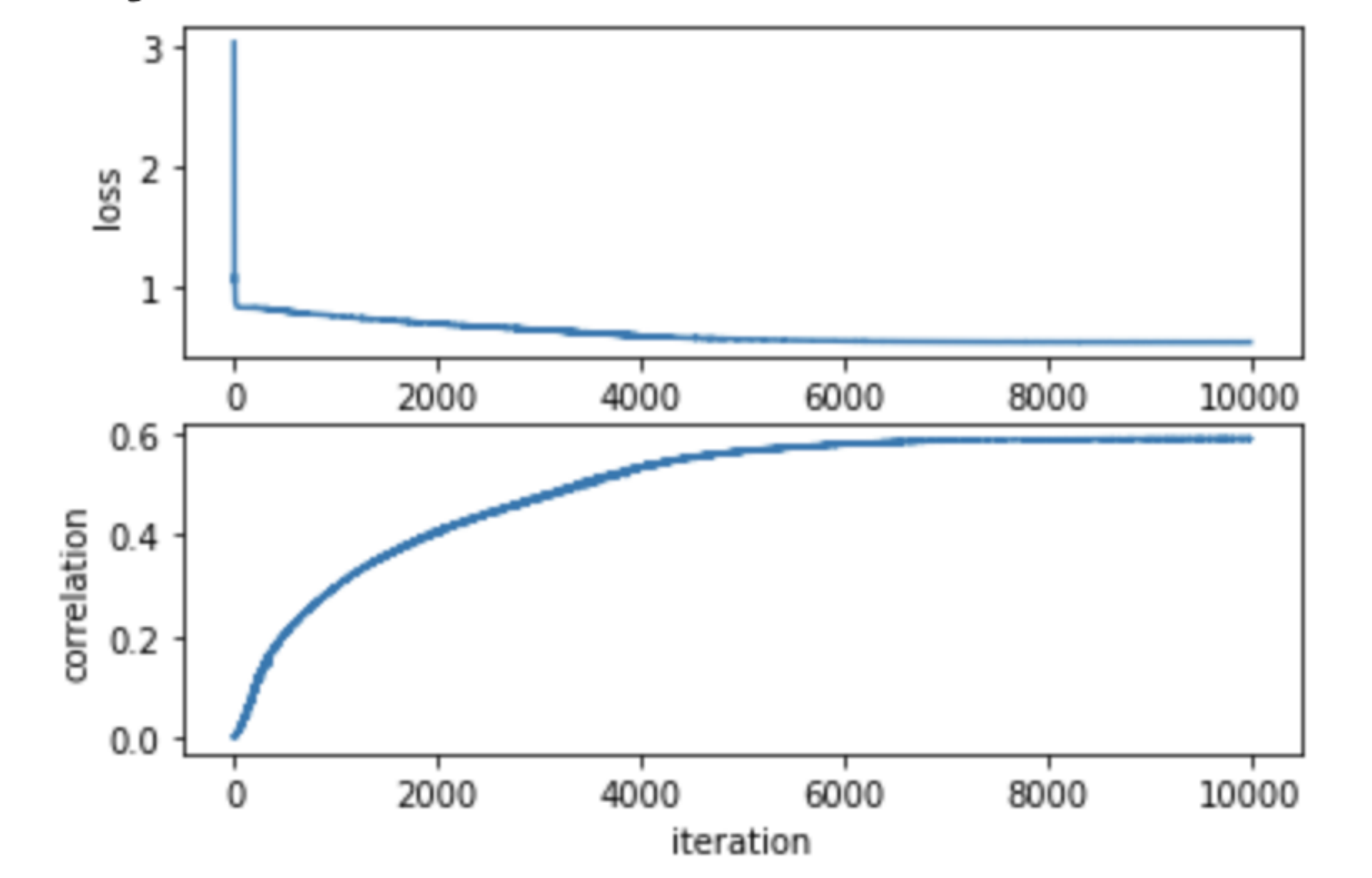}
        \caption{RNN model with AutoEncoder loss curve in training(upper panel), spatial correlation(lower panel)}
        \label{fig:RNN_train}
    \end{subfigure}
    \hfill
    \begin{subfigure}[b]{0.3\textwidth}
        \centering
        \includegraphics[width=\textwidth]{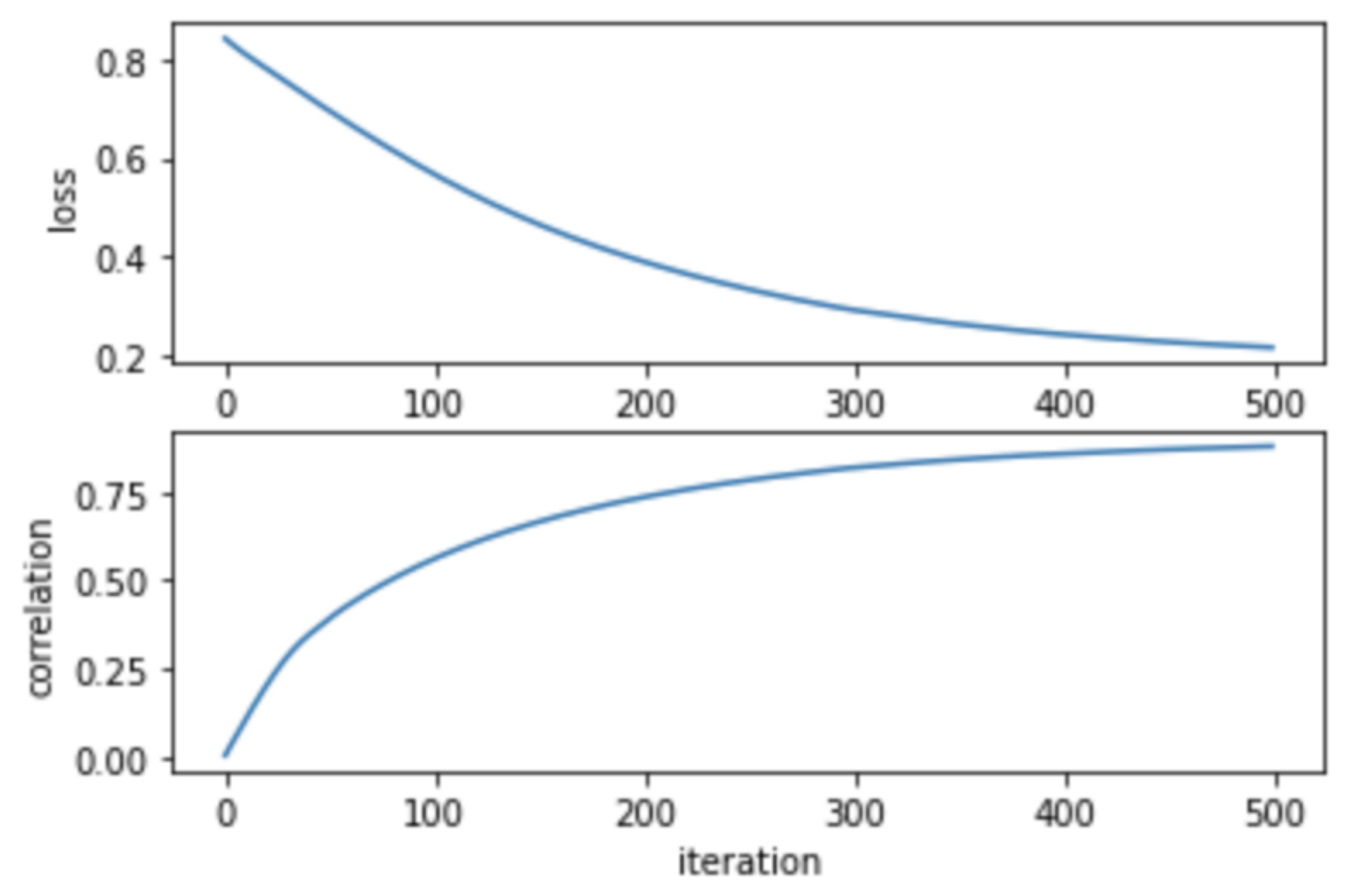}
        \caption{Pure RNN model loss curve in training(upper panel), spatial correlation(lower panel)}
        \label{fig:pure_RNN_train}
    \end{subfigure}
    \caption{Loss curve and spatial correlation in training for three different model architecture. ODE and RNN with AutoEncoder will take $10^4$ epochs to converge and attain 0.6 spatial correlation. Pure RNN converges very fast and will achieve 0.8 spatial correlation for training data in 500 epochs.}
    \label{fig:loss and train correlation}
\end{figure}
The training was run on the GPU server Tesla K80. We selected Adam optimizer with learning rate $10^{-4}$, $\beta_1=0.9$ and $\beta_2=0.99$. We focused our evaluation on training loss, spatial correlation on 2D image data. For training strategy of this network. We first trained the network to learn the spatial latent representation, which can be regarded as AutoEncoder. Then we trained the network to learn the spatial temporal latent code and ODE network. We focused our training on spatial temporal latent code part. It takes $10^4$ epochs to train and the loss cuve is plotted on top of Figure \ref{fig:ODE_train}. On lower panel of the Figure \ref{fig:ODE_train}, it depicts the spatial correlation during training. The final training correlation is  close to 0.6. The training is successful and maintain the average spatial correlation in a high level. After training, we tested on 150 subjects to see the spatial correlation versus time point. We found the spatial correlation for each time point is significantly high enough, and correlation value is around 0.53, which means the spatial temporal latent representation is learned well and ODE can be used as backbone to establish the temporal relationship between different latent code.

\subsection{Compare to other architecture}
We tested our spatial temporal representation learning on other two architectures by replacing the Neural ODE as RNN. The aim is to test whether the latent code is learned successful for different temporal model. One of the model architecture is shown in Appendix \ref{Appendix A} Figure \ref{fig:rnn_w_ae_archi} and training loss curve, spatial correlation at training time is shown in Figure \ref{fig:RNN_train}. We got similar result compared with Neural ODE which means the spatial temporal latent representation is the key in success of video prediction in resting state fMRI data. We also tested a intuitive temporal modeling by using pure RNN to establish temporal relationship between frames of different time points. In pure RNN following architecture in  \cite{hochreiter1997long}, we didn't use spatial temporal latent code but just compute spatial latent code and predict the future frame one by one. We didn't do grouping for the trajectory and treat each frame as the output of the network. The model architecture is listed in Appendix \ref{Appendix A} Figure \ref{fig:rnn_wo_ae_archi}. The model converges in 500 epochs and spatial correlation in training is high around 0.8 as shown in Figure \ref{fig:pure_RNN_train}. However in testing time, the spatial correlation is low and drop very fast versus time. The result is as expected, since neither RNN and ODE can predict very long time sequence with high quality. What we should do is to compress the data not only in spatial but also in temporal to reduce prediction time points in RNN or ODE and improve prediction quality.

\subsection{Forward and bidirectional prediction}
\begin{figure}[H]
    \centering
    \begin{subfigure}[b]{0.3\textwidth}
        \centering
        \includegraphics[width=\textwidth]{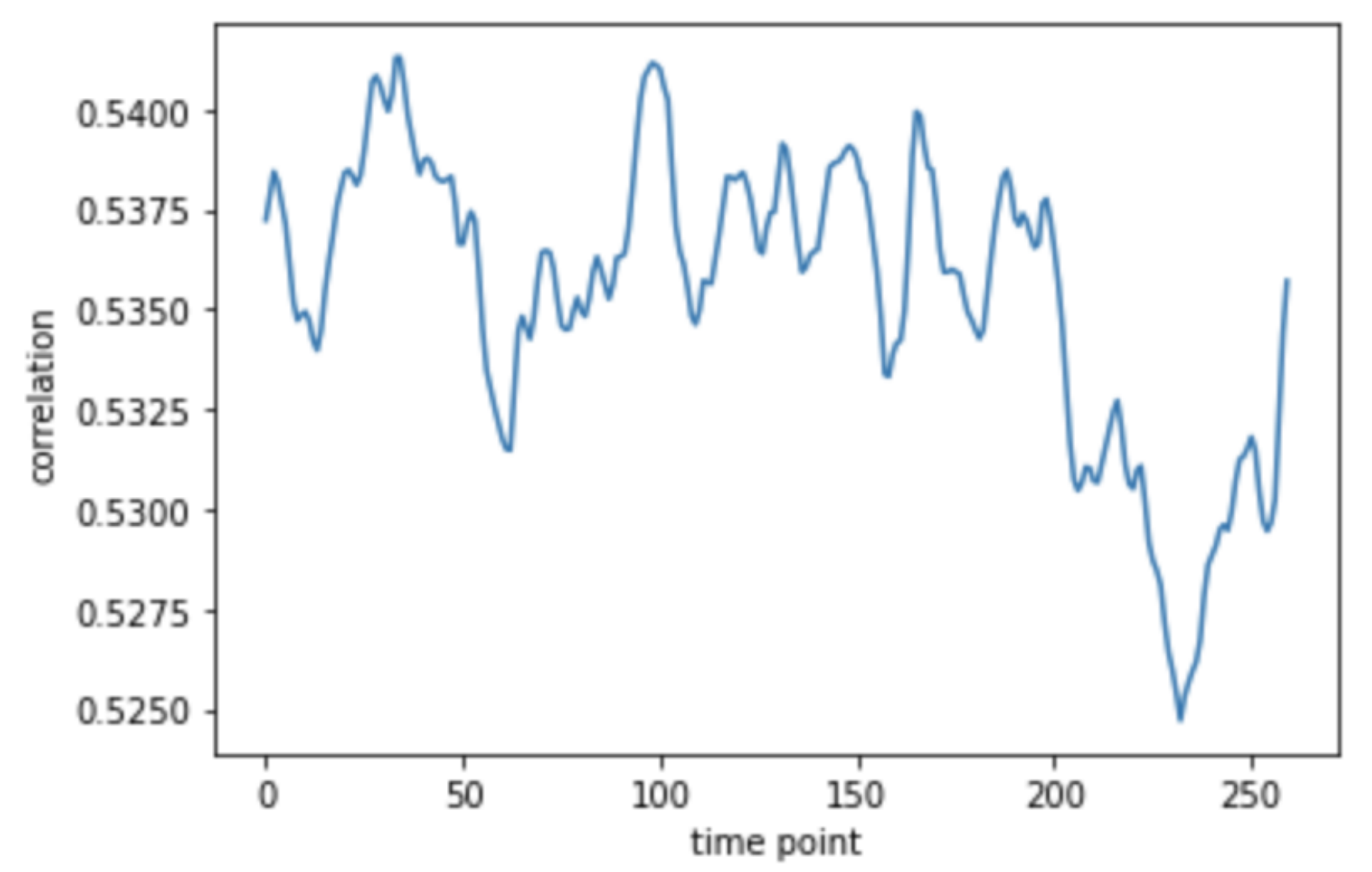}
        \caption{ODE model spatial correlation for test data.}
        \label{fig:ODE test corr}
    \end{subfigure}
    \hfill
    \begin{subfigure}[b]{0.3\textwidth}
        \centering
        \includegraphics[width=\textwidth]{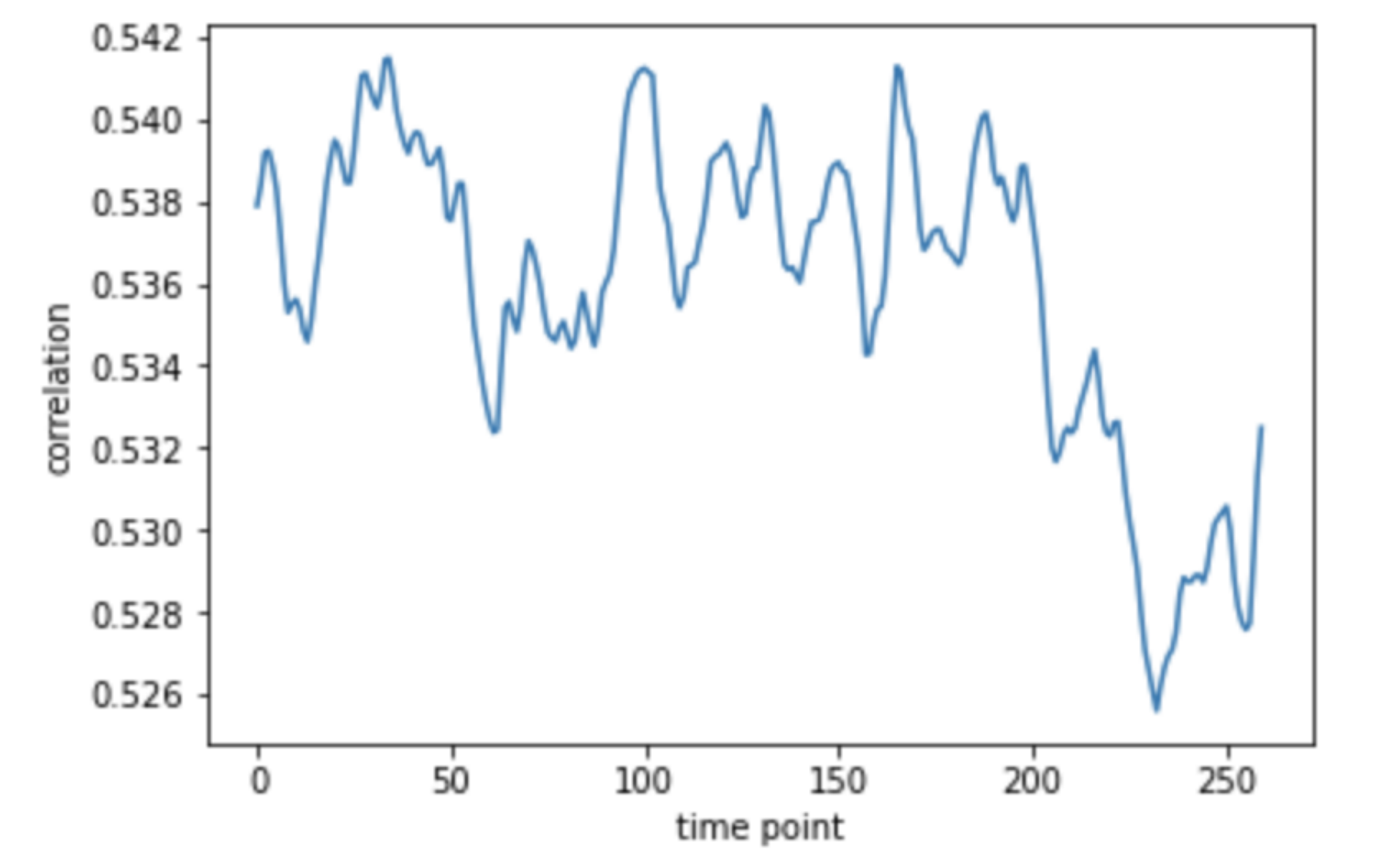}
        \caption{RNN model with AutoEncoder spatial correlation for test data.}
        \label{fig:RNN test corr}
    \end{subfigure}
    \hfill
    \begin{subfigure}[b]{0.3\textwidth}
        \centering
        \includegraphics[width=\textwidth]{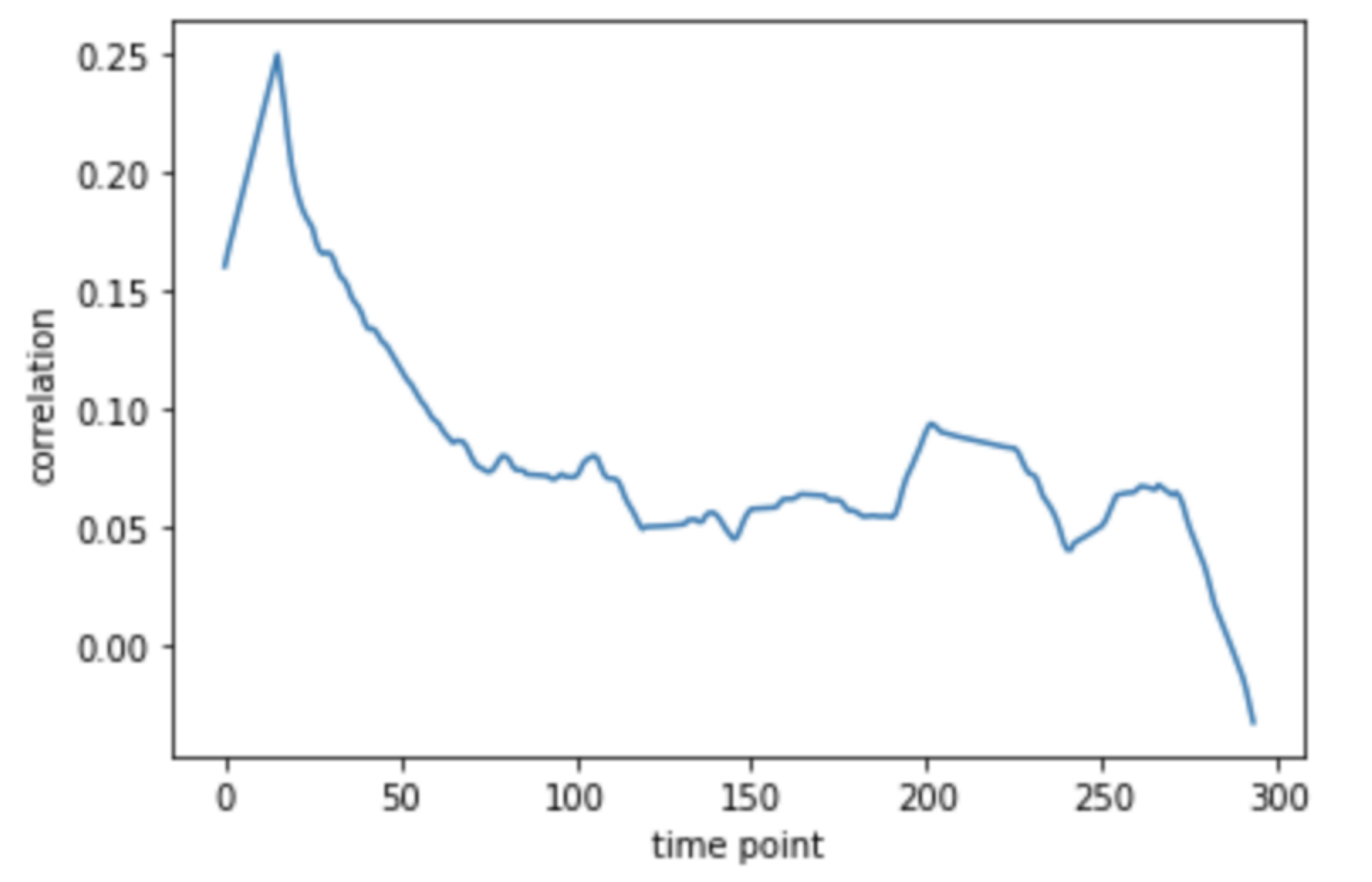}
        \caption{Pure RNN model spatial correlation for test data.}
        \label{fig:pure_RNN test corr}
    \end{subfigure}
    \caption{Spatial Correlation for three different model. The first two model achieve high correlation versus time point, and the performance is stable. Pure RNN model prediction quality drop fast and go to zero after 50 time points prediction.}
    \label{fig:test spatial correlation}
\end{figure}

\begin{figure}[h]
    \centering
    \includegraphics[width=1\textwidth]{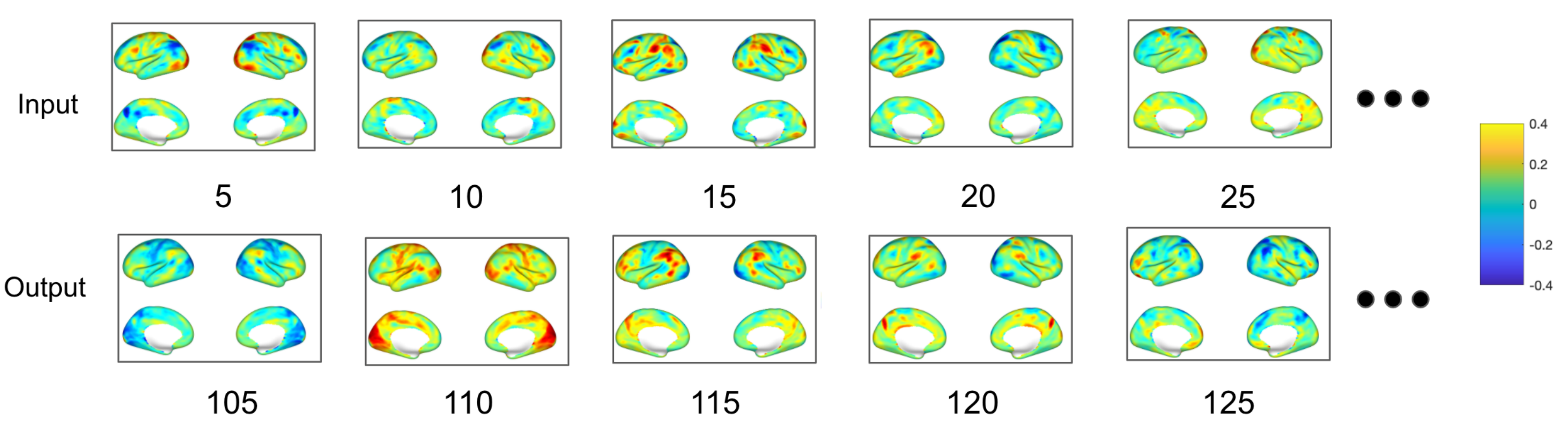}
    \caption{Given the first 100 frames of fMRI data, ODE model help predict left 300 frames for each subject.}
    \label{fig:forward prediction}
\end{figure}

We had 150 subjects data for testing. The test consisted three parts: forward prediction, bidirectional prediction and VAE resampling. The qualitative evaluation of forward prediction can be seen in the Figure \ref{fig:forward prediction}. We had 100 concatenated transformed 2D fMRI data as input. They were compressed into spatial temporal latent code and were propagated forward to predict three subsequent latent codes $z_t$ following the given $z_0$. We decoded these three subsequent codes into 300 2D fMRI data. Here we just show the 5 example input images and 5 example output images. More quantitative evaluation can be see in Figure \ref{fig:test spatial correlation}. The x-axis is output time points and y-axis is spatial correlation. We deleted the first few and last few time point in the first two subfigure since they are lower than average abnormally, which can be left for further study. Figure \ref{fig:ODE test corr} reveals our model performance when ODE is used as backbone for video prediction. Spatial correlation is high throughout the whole prediction trajectory. Figure \ref{fig:RNN test corr} and Figure \ref{fig:pure_RNN test corr} reveal the performance of the trained network shown in Appendix \ref{Appendix A}. The result reflects the success of training and latent code contain useful information for further usage.

\begin{figure}[h]
    \centering
    \includegraphics[width=1\textwidth]{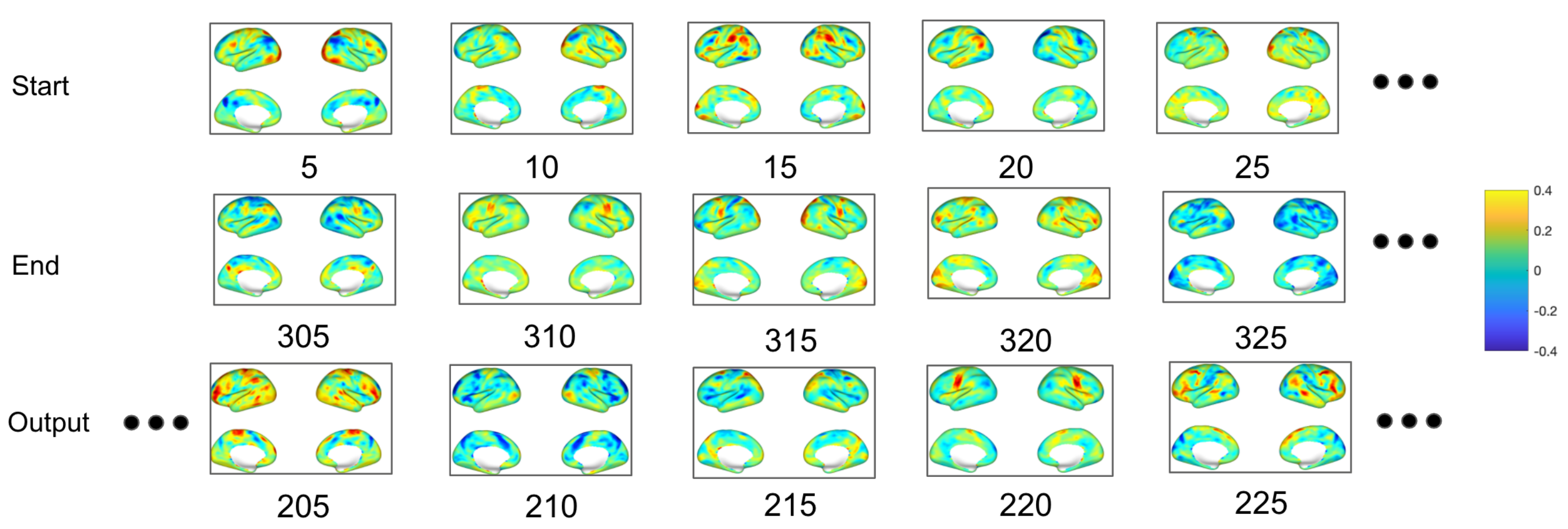}
    \caption{Given the first 100 frames, last 200 frames of fMRI data, ODE model help do interpolation for the left 100 frames for each subject.}
    \label{fig:bidirectional prediction}
\end{figure}

The second test is bidirectional prediction which is trained following Equation \ref{eqn:loss for bidirectional pred}. In test time, we were given the first 100 2D fMRI data. The model was asked to predict the whole trajectory following Equation \ref{eqn:ODE function} and \ref{eqn:integral on ODE}. Figure \ref{fig:bidirectional prediction} plots the training input and prediction output in image space. The first row is first several starting fMRI data and second row is the last few fMRI data in the same trajectory. The output is interpolation of the fMRI data in the middle of 'Start' and 'End'.

\begin{figure}[h]
    \centering
    \includegraphics[width=1\textwidth]{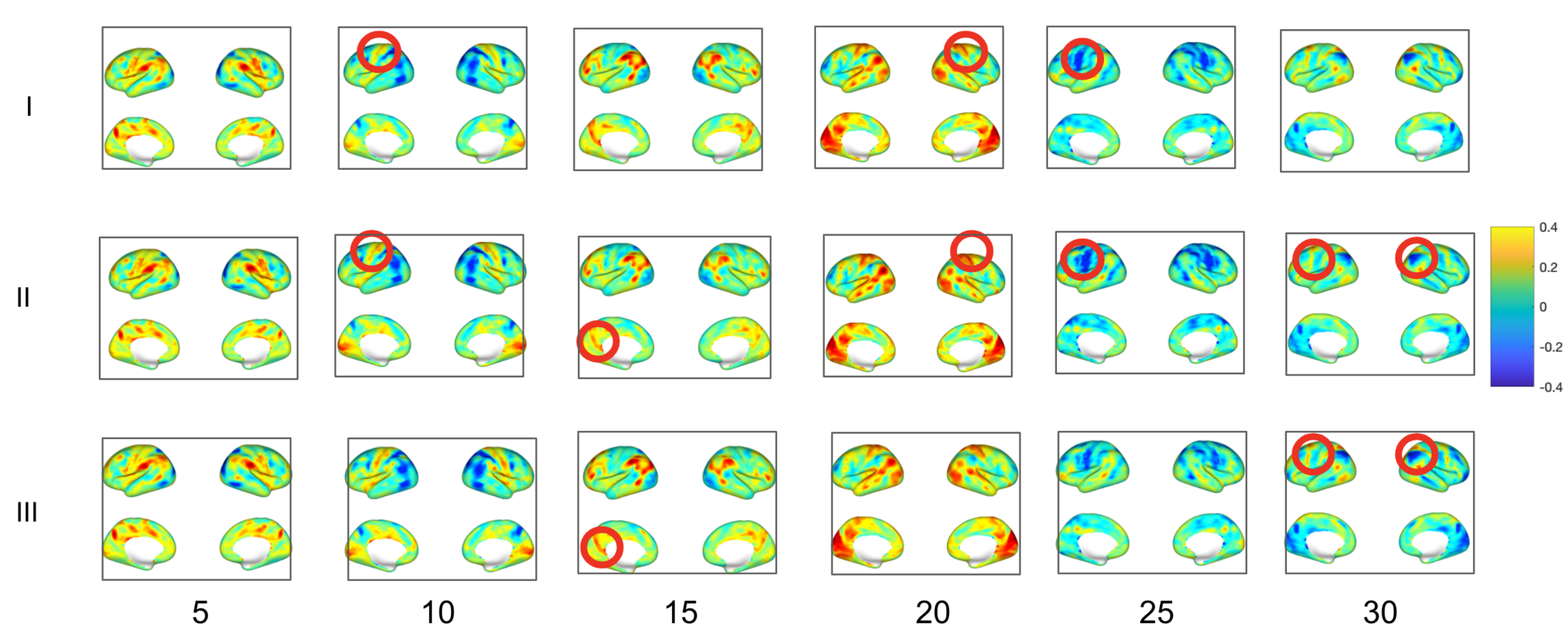}
    \caption{We model the latent representation as prior Gaussian distribution. Different sample will result in image sequence with slight difference. Red circle point out the difference caused by VAE model.}
    \label{fig:vae sample}
\end{figure}

Besides predicting the image and calculate the spatial correlation as evaluation, in Equation \ref{eqn:vae distribution}, \ref{eqn:reparametrize} and \ref{eqn: loss for vae}, we introduced Variational AutoEncoder to explain the variance of the latent code. We tested to see how the latent distribution influence the output. We used the same trajectory as input and test for 100 times. Three output trajectories were selected to compare their difference. We show the result in Figure \ref{fig:vae sample}. They are basically the same but differ a little bit in some particular region. The regions with difference are pointed out by red circle.

\subsection{Three cluster center best explain GMM model}

\begin{figure}[h]
    \centering
    \begin{subfigure}[b]{0.35\textwidth}
        \centering
        \includegraphics[width=1\textwidth]{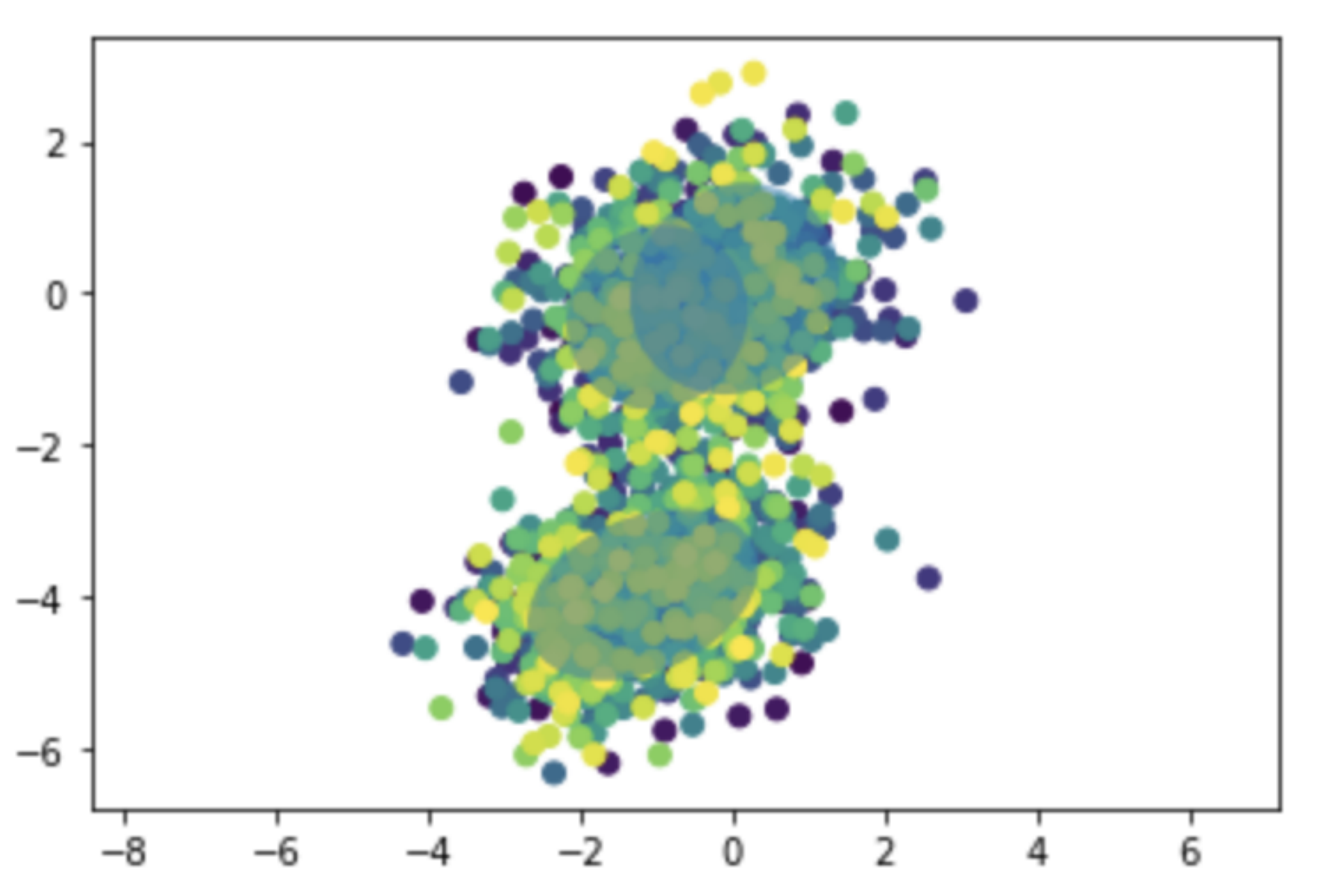}
        \caption{Spatial temporal latent representation distribution in first two dimensions. Ellipsoids depict mixture of three Gaussian distribution.}
        \label{fig:Latent code distribution}
    \end{subfigure}
    \begin{subfigure}[b]{0.35\textwidth}
        \centering
        \includegraphics[width=1\textwidth]{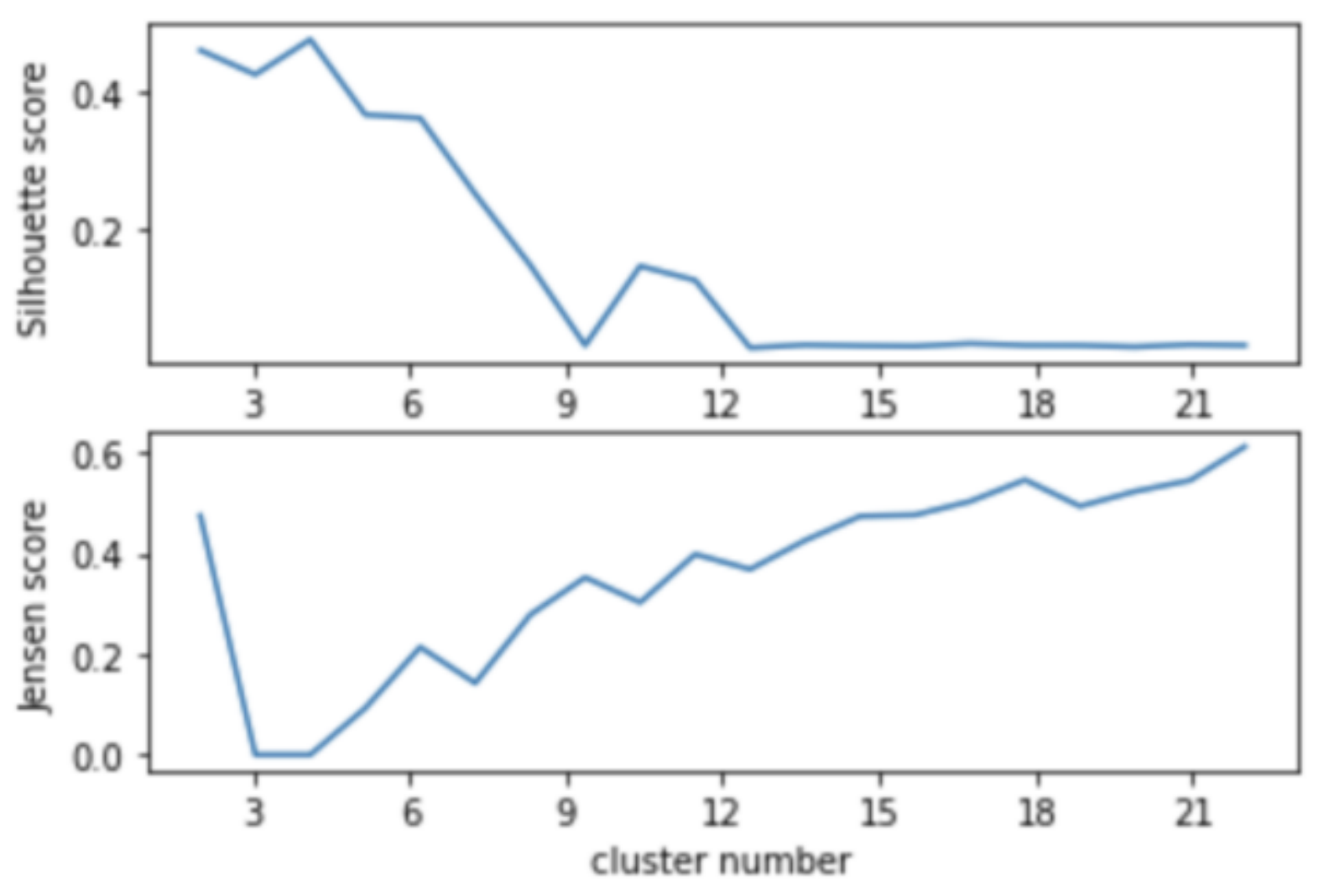}
        \caption{Silhouette score and Jensen score for selecting cluster number. High Silhouette score and low Jensen score is preferred.}
        \label{fig:Two score}
    \end{subfigure}
    \par
    \begin{subfigure}[b]{0.8\textwidth}
        \centering
        \includegraphics[width=1\textwidth]{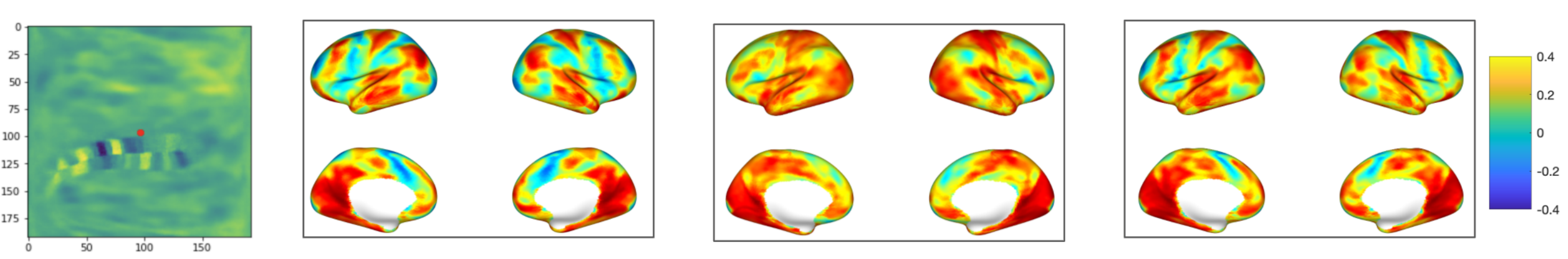}
        \caption{Three seed based temporal correlation for three cluster centers shown in (a). Seed point is shown as red point in first subfigure.}
        \label{fig:temporal correlation for cluster center.}
    \end{subfigure}
    \caption{Gaussian Mixture Model is performed in latent space for 150 subjects. We select cluster number that Silhouette score is high while Jensen score is low. Each cluster center is decoded into image space corresponding to 100 image frames each. Temporal correlation is computed to reveal the common pattern.}
    \label{fig:GMM analysis}
\end{figure}

In section 3.3, we used Gaussian Mixture model to describe the distribution of spatial temporal latent code. To select a suitable number of cluster, we computed two metrics which are Silhouette score and Jensen Shannon score as criteria. Silhouette score is a method to validate consistency within clusters of data. The score ranges from $-1$ to $+1$, where a high value indicates that the high dimensional latent code in each cluster is well assigned, and the clustering configuration is appropriate. The Jensen Shannon score is a method of measuring the similarity between two probability distributions. It is also known as total divergence to the average. We had to select a cluster number that minimize the Jensen Shannon score. As shown in Figure \ref{fig:Two score}, when cluster number is 3, Silhouette score is high while Jensen Shannon score is low. In Figure \ref{fig:Latent code distribution}, we plot the latent code distribution and cluster ellipsoid in the first two dimension of high dimensional latent code.

To further explore the property of three cluster centers, we first decoded the cluster center into three groups of images with each group 100 fMRI data frames. Then we computed the seed based temporal correlation and plot in Figure \ref{fig:temporal correlation for cluster center.}. Seed location is plotted as a red point in the first subplot, and followed by three temporal correlation map in image space. The first subplot is 2D images representing surface of the brain map that we extract from HCP Volumetric data following \cite{kim2020representation}
\subsection{Latent code change along gradient flow}
\begin{figure}[h]
    \centering
    \includegraphics[width=1\textwidth]{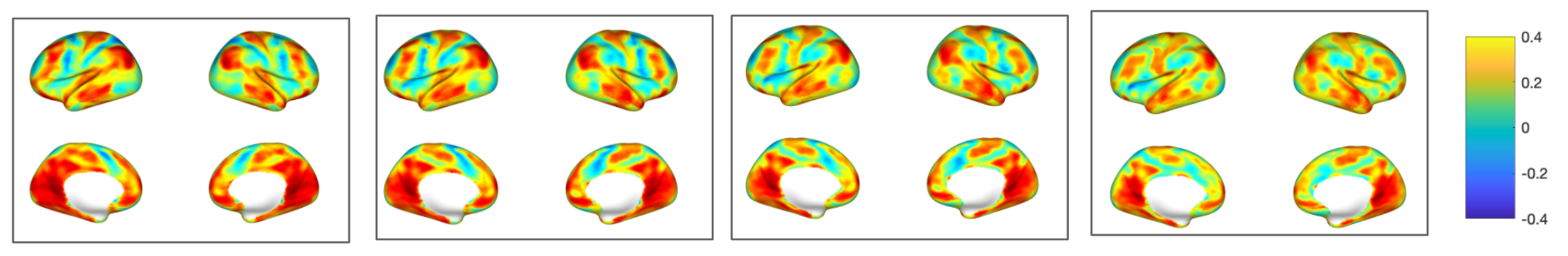}
    \caption{We select the equilibium point of the system and let latent code go along one gradient direction. Then we decode the latent representation to groups of images. Temporal correlation is calculated to reveal change of image space correlation value versus the change in latent space.}
    \label{fig:gradient flow image space.}
\end{figure}

In section 3.6, we talked about the analysis on trained Neural ODE parameter. After training, we can compute the equilibrium of ordinary differential equation. The equilibrium is also a latent code. We want to know how the change of latent code lead the change of temporal correlation for groups of images. The latent code was changed along the gradient flow by a predefined step size, and we obtained 4 different latent codes along this gradient flow starting from equilibrium. The latent code was decoded into groups of images. Seed based temporal correlation was calculated for these four groups of images separately and are depicted in Figure \ref{fig:gradient flow image space.}. The temporal correlation map shows the decreasing correlation of other area against the seed we selected. This may help understand how the brain pattern change on the direction of gradient flow provided by the trained ODE model.
\subsection{Human traits analysis using latent space}
The spatial temporal latent code can be further used to estimate human traits for each subject in HCP data. We selected six human traits including passive traits and active tasks as target for our prediction. The value of each trait can be obtained accompanying with the record of fMRI data. We added one linear layer with latent code as input and output the estimation of different human traits. Then we calculate the error of prediction for these 6 different human traits by Normalized Root Mean Squared Error(NRMSE). The result is summarized in the table \ref{tab:table}.

\begin{table}[H]
 \caption{Human traits prediction error}
  \centering
  \begin{tabular}{lllllll}
    \toprule
    Name     &FearAffect & Sadness &LifeSatif &Friendship &LanguageTask Acc & RelationalTask Acc\\
    \midrule
    NRMSE & 0.016 & 0.024 & 0.014 & 0.012 & 0.008 & 0.013\\
    \bottomrule
  \end{tabular}
  \label{tab:table}
\end{table}
Our latent code can provide a very accurate prediction of the human traits for different subjects which means characteristic of fMRI trajectory for different subjects can be successfully revealed by latent code. There are very deterministic relationship between the trajectory information and human behavior.
\section{Discussion}
In this paper, we proposed a new method of video frame prediction for resting state fMRI data and made use of latent code to do some downstream tasks. The key point for success is the compression on both spatial and temporal information. The difference between Neural ODE and RNN is that in training time, Neural ODE is a recursive forward propagation method and output of first ODE block will be the input of the next ODE block. RNN is always used to do a one step forward prediction in training while in testing, it generate prediction recursively. Either ODE or RNN will work well regarding our spatial temporal latent code and differ little in performance. However ODE can be easily used to interpret how the latent code in one time stamp transferred to the next by analyzing differential equation while RNN lack this interpretability.

Besides the prediction quality, the geometry explanation of latent space and utilization of ODE trained parameters are two main concern in this paper. \cite{kim2020representation} explained geometry of spatial representation of resting state fMRI as a sphere in high dimension which differs for different subjects. While in forward prediction in our temporal dynamic model, the latent space would not necessarily be Gaussian distribution and would be influenced by both starting point of forward prediction and ODE trained parameter. In experiment, we analyzed the distribution of latent code of ODE prediction. The visualization of first two variables in latent code reveals that it is not Gaussian but can be explained by Gaussian mixture. According to Silhouette score and Jensen Shannon score, it revealed that three Gaussian Mixture can explain well to the resulting latent code distribution. The latent code was transferred from zero mean Gaussian to another Gaussian with different mean value, but the variance change little, which is due to the combination of prior Gaussian assumption for $z_0$ and ODE propagation for $z_t$.

We tested the trained model for forward prediction, bidirectional prediction and sampling from VAE latent space to get qualitative evaluation. The prediction looks good and VAE also explain the variance of the latent code and show in image space as shown in Figure \ref{fig:vae sample}. The region within the red circle may reveal the subject difference among whole population.

Another potential usage of this model in fMRI data is to do interpolation for smaller group of fMRI images. The measurement of fMRI data always include some random noise and physiological noise. \cite{power2014methods} proposed censoring method to eliminate the head motion in fMRI data that is significant. \cite{laumann2017stability} used censoring method to help understand time varying temporal correlation but that method is criticized for it destroys the temporal relationship. But ODE can interpolate the censored data if we know which time points are censored. In work of \cite{khazaee2017classification} and \cite{du2018classification}, the author used fMRI data to do diagnosis. Our spatial temporal latent code can also be used in disease diagnosis and human traits prediction as shown in section 4.7.

Overall, we treat the temporal dynamic modeling for fMRI data as a video prediction problem. The model accurately predicts the image frames given a group of images. Analysis on latent space and image space may shed light on the study of temporal dynamic of resting state fMRI data.

\bibliographystyle{plainnat}
\bibliography{references}  

\newpage
\appendix
\section{Network architecture of RNN w/ and w/o AE}
\label{Appendix A}
\begin{figure}[H]
    \centering
    \begin{subfigure}[b]{0.45\textwidth}
        \centering
        \includegraphics[width=1\textwidth]{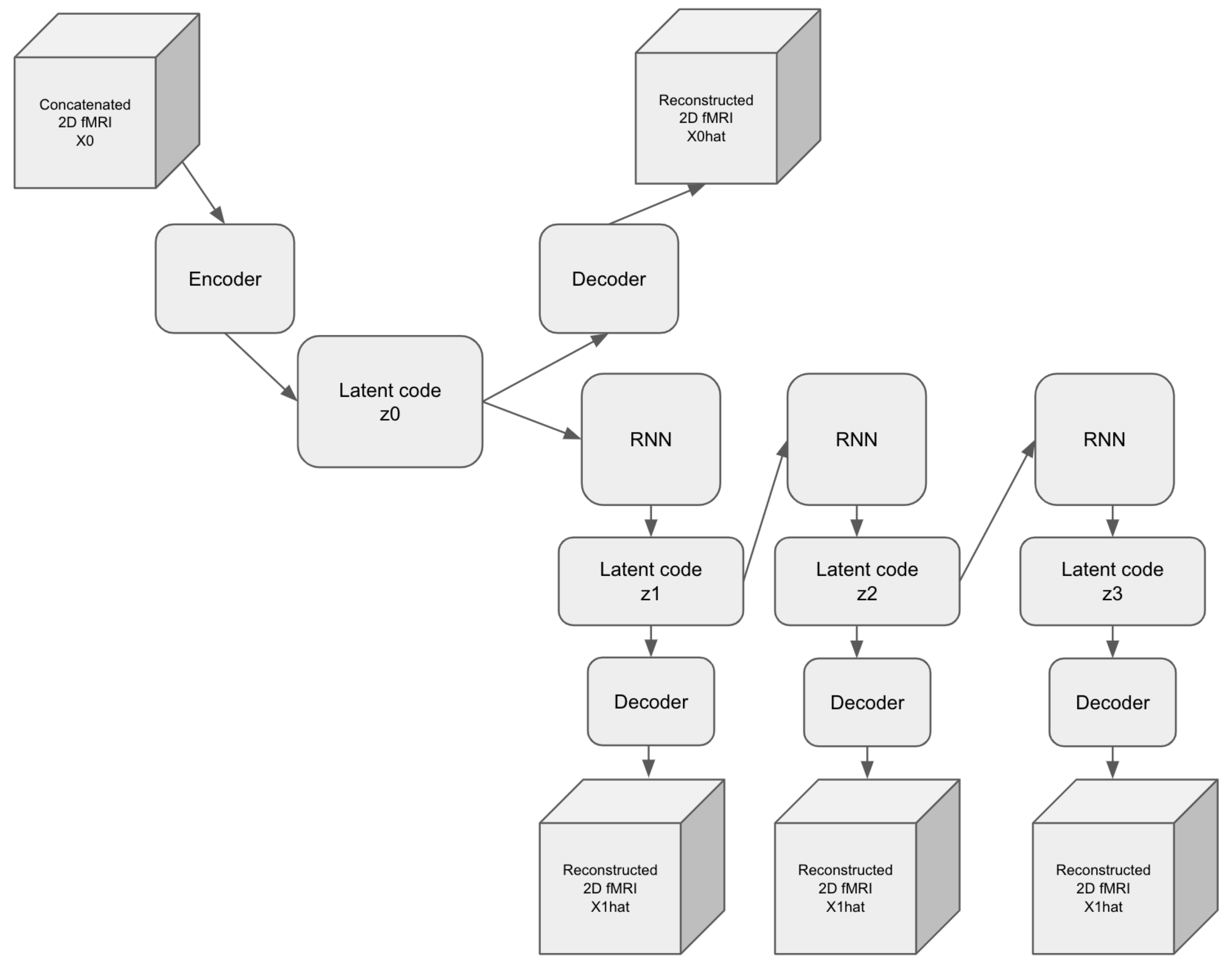}
        \caption{ODE is replaced by RNN to predict the spatial temporal latent representation. Output of first RNN will be the input of the next RNN. Latent code is decoded to recover the data in image space.}
        \label{fig:rnn_w_ae_archi}
    \end{subfigure}
    \hfill
    \begin{subfigure}[b]{0.45\textwidth}
        \centering
        \includegraphics[width=1\textwidth]{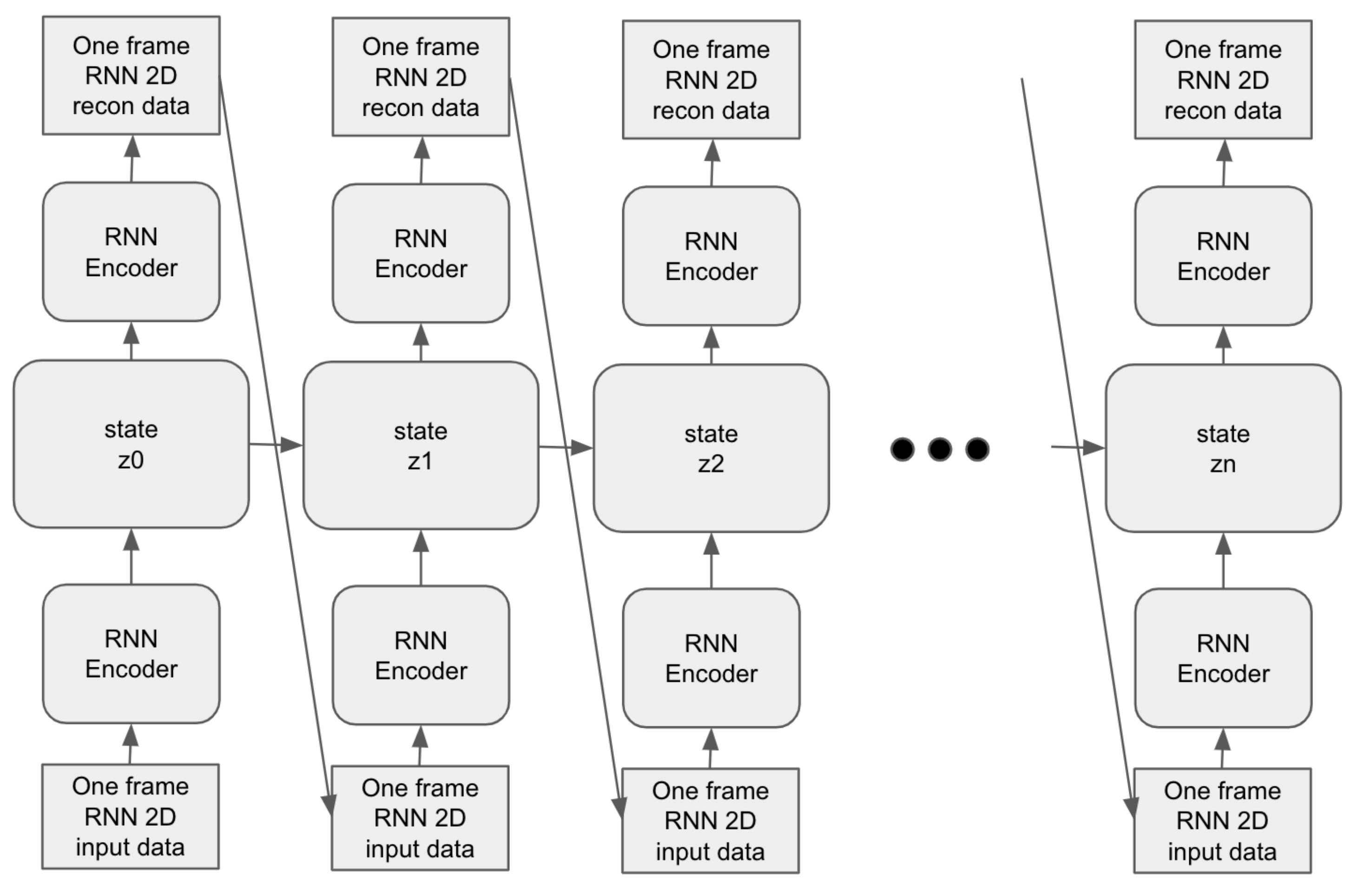}
        \caption{Traditional RNN architecture to predict spatial latent code. The output of RNN represents only one time points rather than groups of images. In test time, the prediction is recursively generated.}
        \label{fig:rnn_wo_ae_archi}
    \end{subfigure}
    \caption{We tried two other architecture for forward prediction of fMRI data. The key success of high quality prediction is do compression on data not only in spatial space but also in temporal space. The architecture on the right can predict one step forward very accurate but performance drop quickly.}
    \label{fig:two other architecture}
\end{figure}

\section{EM algorithm in GMM}
\label{Appendix B}
We initialized means $\mu_k$, covariances $\Sigma_k$, and mixing coefficients $\pi_k$ for K Gaussians in Equation \ref{eqn:gmm max prior dist}.\\
E step:\\
\begin{equation}
    \label{eqn:E step parameter update}
    \gamma(s_{tk})=\frac{\pi_k\mathcal{N}(\Vec{x}_t|\mu_k, \Sigma_k)}{\sum_{j=1}^K\pi_j\mathcal{N}(\Vec{x}_t|\mu_j, \Sigma_j)}
\end{equation}
M step:
\begin{align}
    \label{eqn:M step parameter update}
    &\pi_k^{new} = \frac{N_k}{N}=\frac{\sum_n\gamma(s_{tk})}{N}\\
    &\mu_k^{new}=\frac{1}{N_k}\sum_{n=1}^N \gamma(s_{tk})z_t\\
    &\Sigma_k^{new} = \frac{1}{N_k}\sum_{n=1}^N \gamma(s_{tk})(z_t-\mu_k^{new})(z_t-\mu_k^{new})^T
\end{align}
After several steps of EM, we could compute the mean and variance for each cluster. We decoded the clustering center to analyze the common brain patterns embedded in spatial temporal latent code.

\end{document}